\documentclass[aps,pra,amsmath,amssymb,showpacs,twocolumn]{revtex4}

\usepackage[T1]{fontenc}
\usepackage{amssymb,amsmath}
\usepackage{amsfonts}
\usepackage{graphicx}
\usepackage{amsthm}
\usepackage{color}
\usepackage{mathbbol}

\def\ket#1{|#1\rangle}
\def\ketbra#1{|#1\rangle\langle #1|}

\def\bra#1{\langle#1|}

\def\pt{\mathrm{PT}}
\def\ppt{\mathrm{PT}}
\def\tr{\mathrm{tr}}
\def\bfa{\mathbf{a}}
\def\bfb{\mathbf{b}}
\def\bfi{\mathbf{i}}

\def\bfz{\mathbf{0}}

\def\cals{\mathcal{S}}
\def\Rr{R^{(r)}}

\def\matrixT{T}
\def\matrixTr{T^{(r)}}

\newcommand{\bfmu}{\boldsymbol{\mu}}
\newcommand{\bfnu}{\boldsymbol{\nu}}
\newcommand{\bftau}{\boldsymbol{\tau}}

\newcommand{\cohalpha}{\alpha}

\newcommand{\cH}{{\cal H}}

\newtheorem{theorem}{Theorem}[]

\begin{document}

\title{Partial transpose criteria for symmetric states}


\author{F.~Bohnet-Waldraff$^{1,2}$, D.~Braun$^{1}$, and O.~Giraud$^{2}$}
\affiliation{$^{1}$Institut f\"ur theoretische Physik, Universit\"at T\"{u}bingen,
72076 T\"ubingen, Germany\\
$^{2}$\mbox{LPTMS, CNRS, Univ.~Paris-Sud, Universit\'e Paris-Saclay, 91405 Orsay, France}
}

\begin{abstract}
We express the positive partial transpose (PPT) separability criterion
for symmetric states of multi-qubit systems in terms of matrix
inequalities based 
on the recently introduced tensor representation for spin states. We construct a matrix from the tensor representation of the state and show that it is similar to the partial 
transpose of the density matrix written in the computational
basis. Furthermore, the positivity of this matrix is equivalent to the
positivity of a correlation matrix constructed from tensor products of
Pauli operators. 
This allows for a more transparent experimental interpretation of the PPT criteria for an arbitrary spin-j state. The unitary matrices connecting our matrix to the partial transpose of the state generalize the so-called magic basis that plays a central role in Wootters' explicit formula for the concurrence of a 2-qubit system and the Bell bases used for the teleportation of a one or two-qubit state.
\end{abstract}
\date{June 24, 2016}
\maketitle

\section{Introduction}
\label{sec:intro}
Quantum information provides a window on various remarkable features
of quantum mechanics, such as entanglement \cite{4Hor09} or
teleportation \cite{BenBraCre93}.  A central resource in quantum
information processing is quantum entanglement. A quantum
state is said to be separable if it can be written as a convex sum of
product states, i.e.~states that are tensor products of states
of all the subsystems; otherwise it is said to be entangled
\cite{Werner89}. The state 
of a bipartite quantum system is known to be separable iff it remains
positive under all positive quantum maps.  
Looking at a subclass
of positive quantum channels, one obtains
necessary conditions for separability, which therefore signal
bipartite entanglement if the condition is violated. In this respect, a central role is played by the
``positive-partial-transposed criterion'' (PPT), physically 
obtained by time-reversal of one of the two
subsystems \cite{Peres96,Horo3}. For systems with Hilbert-space
dimensions at most $2\times 2$ or 
$2\times 3$, PPT is also sufficient for separability. For higher
dimensional systems, entangled states exist that have
positive partial transpose
\cite{horodecki_separability_1997,amselem_experimental_2009}.

For multipartite systems, the 
situation is substantially more complicated due to the possibility that only certain bi-partitions be entangled \cite{dur_classification_2000}. For three qubits, six different
SLOCC-equivalence classes exist (i.e.~families of states that can be transformed
into each other with non-zero probability using only stochastic local
operations and classical 
communication), including two of genuine multipartite entanglement
\cite{dur_three_2000,Acin2001}. For four qubits, there are already uncountably  
many SLOCC-classes \cite{gour_all_2010}.  Polynomial invariants
(under SLOCC)
have been used to classify and even quantify the entanglement of
multi-qudit states \cite{viehmann_polynomial_2011,gour_classification_2013}.\\ 

For symmetric states, that is, states belonging to the vector space spanned by pure states invariant under particle exchange, the situation is somewhat simpler, in the sense
that several entanglement criteria coincide \cite{TothGuehne}. Continuous sets of SLOCC-classes of pure states can be grouped into SLOCC-invariant families based on the degeneracy structure of the involved single particle states \cite{bastin_operational_2009}.  Notably, PPT is equivalent to the positivity of a correlation matrix of moments of local orthogonal observables \cite{TothGuehne}.   PPT symmetric states of two or three qubits are all separable \cite{eckert_quantum_2002}, whereas for four, five, or six qubits entangled symmetric  PPT states exist \cite{TurLew12,Toth09}.  PPT mixed symmetric states for $N$ qubits were studied in \cite{augusiak_entangled_2012}, where criteria for separability in terms of the ranks of such states were found. 

In a parallel line of research, the concept of classical spin states
and the notion of quantumness of a spin state was introduced
\cite{GirBraBra08,Giraud10,Martin10,fabian1}. In analogy to quantum
optics, a pure spin state is considered 
(most) classical if the quantum fluctuations of the spin vector are
minimal, i.e.~as small as allowed by Heisenberg's uncertainty
principle. This selects uniquely the SU(2)-coherent states as pure
``classical spin states''. Their convex hull is 
the set of all classical spin states, and the distance of a given
state $\rho$ from this convex set is a measure of its ``quantumness''. In \cite{Giraud10, Martin10}
the quantumness based on the Hilbert-Schmidt distance and the Bures
distance was analyzed, and the ``most quantum'' state for these
measures identified. Classical states of a spin-$j$ are in fact formally identical to fully separable symmetric states of $N=2j$ spins-$\frac12$ (see Section \ref{classvssep}). Statements about the classicality of
spin-$j$ states therefore immediately translate to statements about
the separability of symmetric states of multi-qubit systems. 

In \cite{fabian1}, it was noted that PPT for spin-1 states is
equivalent to the positivity of a matrix built from tensor
entries of a recently introduced tensor representation of the state
\cite{PRL2015}. The aim of the present work is to generalize this
result to 
arbitrary spin and bipartition.  We show
that an appropriate arrangement of the components of the tensor
representing a spin-$j$ state leads to a matrix that is similar to
the partial transposed multi-qubit state written in the computational
basis. Hence, positivity of this matrix is equivalent to PPT of the
multi-qubit state.  We explicitly construct the unitary transformations that
connect the two matrix representations, and show that they generalize
the ``magic basis'' that for two qubits allows one to obtain an
explicit form of the concurrence \cite{wootters_entanglement_1998}. We
also point out 
the connection to correlation functions that were introduced earlier
for studying entanglement of symmetric spin states
\cite{UshPraRaj07,usha_devi_constraints_2007,TothGuehne}. 
After recalling the basic definitions in Section \ref{sec:classical}, we first consider the easier case of an equal bipartition in Section
\ref{evenPPT}, then move on to the general case in Section
\ref{generalPPT}. In Section \ref{applications} we discuss various
consequences of our results.

\section{Classical spin states}
\label{sec:classical}

\subsection{Tensor representation}
Let $\rho$ be a spin-$j$ state (mixed or pure), with $j$ integer or
half-integer, and $N=2j$. In \cite{PRL2015} we introduced a tensorial
representation of $\rho$ as 
\begin{equation}
\label{rhoarbitrary}
\rho=\frac{1}{2^N}\,X_{\mu_1\mu_2\ldots\mu_N} S_{\mu_1\mu_2\ldots\mu_{N}},
\end{equation}
where 
\begin{equation}
\label{coeffdecompPauli}
X_{\mu_1\mu_2\ldots\mu_N} = \tr(\rho\,S_{\mu_1\mu_2\ldots\mu_{N}})
\end{equation}
is a real symmetric tensor (we use Einstein sum convention, summing over repeated
indices). The matrices $S_{\mu_1\mu_2\ldots\mu_{N}}$
can be obtained from the expansion of the matrix corresponding to the $(j,0)$ 
representation of a Lorentz boost along a 4-vector. Alternately, they can be constructed from Pauli matrices $\sigma^{\mu}, 0\leq \mu\leq 3$, with $\sigma_0$ the $2\times 2$ identity matrix,
as the projection of the tensor product
$\sigma^{\mu_1}\otimes\sigma^{\mu_2}\otimes\ldots\otimes\sigma^{\mu_N}$
onto the subspace spanned by pure states invariant under permutation
\cite{PRL2015}. The tensor representation is such that 
\begin{equation}
\label{propx}
\sum_{a=1}^{3}X_{aa\mu_3\ldots\mu_N} = X_{00\mu_3\ldots\mu_N}
\end{equation}
for arbitrary $0\leq\mu_3,\ldots,\mu_N\leq 3$. The matrix $S_{0\ldots 0}$ is the $(N+1)\times (N+1)$ identity matrix, so that in particular the condition $\tr\rho=1$ is equivalent to $X_{0\ldots 0}=1$.

\subsection{Classical states}
A spin-$j$ coherent state $\ket{\alpha^j}$ associated with the Bloch vector $\bf{n}=(\sin\theta \cos\phi,\sin\theta \sin\phi,\cos\theta)$ is defined as
\begin{equation}
\label{spincoherent}
\ket{\alpha^j} =\!\!\! \sum_{m=-j}^j \sqrt{\binom{2j}{j+m}} \left(\cos\frac{\theta}{2}\right)^{j+m}\left(\sin\frac{\theta}{2}e^{-i\phi}\right)^{j-m}\!\!\!\!\!\! \ket{j,m},
\end{equation}
where $\{\ket{j,m}; -j\leq m\leq j\}$ is the usual angular momentum
basis. Such a state has tensor entries given by
$X_{\mu_1\mu_2\ldots\mu_N} =n_{\mu_1}n_{\mu_2}\ldots n_{\mu_{N}}$,
with $n=(1,\bf{n})$ \cite{PRL2015}. For spin-$\frac12$ we denote coherent states
simply by $\ket{\alpha}$. If  $\ket{\alpha^j}$ is written in the
$N$-spins-$\frac12$ computational basis, we have the identities
\begin{equation}
\label{alphajalpha}
\ket{\alpha^j}=\ket{\alpha}\otimes\ket{\alpha}\otimes\ldots\otimes\ket{\alpha},
\end{equation}
the tensor product of $N$ copies of the spin-$\frac12$ coherent state, and
\begin{equation}
\label{coorcoh}
n_\mu=\bra{\alpha}\sigma^\mu \ket{\alpha}.
\end{equation}
In \cite{GirBraBra08} we introduced classical spin states as the convex hull of coherent states, that is, states $\rho$ such that there exists a positive function $P(\cohalpha)$ defined on the unit sphere and verifying
\begin{equation}
\label{rhoascoh}
\rho=\int d\cohalpha P(\cohalpha)\ket{\cohalpha^j}\bra{\cohalpha^j}.
\end{equation}
In tensor terms, classical states are states $\rho$ whose tensor representation is given by
\begin{equation}
\label{rhoascohtens}
X_{\mu_1\mu_2\ldots\mu_N}=\int_{\mathbb{S}} dn P(n)n_{\mu_1}n_{\mu_2}\ldots n_{\mu_{N}},
\end{equation}
where $\mathbb{S}$ is the unit sphere of $\mathbb{R}^3$ and $dn$ is the flat measure on the sphere. Since the we are considering finite-dimensional Hilbert spaces, Caratheodory's theorem ensures that the integral in \eqref{rhoascohtens} can be replaced by a finite sum, so that there exist weights $w_i\geq 0$ and vectors $n^{(i)}=(1,\bf{n^{(i)}})$ such that 
\begin{equation}
\label{xsep}
X_{\mu_1\mu_2\ldots\mu_N}=\sum_i w_i n^{(i)}_{\mu_1} n^{(i)}_{\mu_2} \ldots n^{(i)}_{\mu_{N}}.
\end{equation}

\subsection{Classicality and separability}
\label{classvssep}
A spin-$j$ state can be seen as the projection of the state of $N$
spins-$\frac12$ onto the vector space $\mathcal{S}$ spanned by pure symmetric states. We
call a mixed state $\rho$ {\it symmetric} if it is equal to its
projection onto $\mathcal{S}$. If a convex combination of pure states
$\rho=\sum w_i \ketbra{v_i}$, with $\ket{v_i}$ pure states and $w_i
\geq 0$, is symmetric, then necessarily all $\ket{v_i}$ belong to
$\mathcal{S}$. Indeed, let $\cals^{\perp}$ be the vector space
orthogonal to $\mathcal{S}$. Then for any vector
$\ket{u}\in\cals^{\perp}$ the symmetry of $\rho$ implies that
$\bra{u}\rho\ket{u}=0$, thus $\sum w_i
|\bra{u}v_i\rangle|^2=0$. Positivity of the $w_i$ then implies that
$\bra{u}v_i\rangle=0$, and thus $\ket{v_i}\in
(\cals^{\perp})^{\perp}=\cals$.

Classical spin-$j$ states can thus be seen as separable fully
symmetric states of $2j$ spins-$\frac12$ states, and vice versa, via the
following theorem: 
\begin{theorem}
\label{th1}
A symmetric state is (fully) separable iff there exists a
P-representation for which the $P$-function is positive on the
two-sphere. In other words, classical states are identified with fully
separable symmetric states. 
\end{theorem}
This theorem was proved many times in many guises (see e.g.~\cite{KorCirLew05} p.4 or \cite{UshRajKar13}). For completeness we briefly give a proof of this fact.
\begin{proof}
If $\rho$ is fully separable, then it is possible to write $\rho=\sum_i\lambda_i\rho^{(i)}_1\otimes\ldots\otimes \rho^{(i)}_N$, and then to decompose each $\rho^{(i)}_k$ in its eigenvector basis, so that 
\begin{eqnarray}
\rho&=&\sum_i\mu_i\ketbra{v^{(i)}_1}\otimes\ldots\otimes \ketbra{v^{(i)}_N}\nonumber\\
&=&\sum_i\mu_i\ketbra{v^{(i)}_1\ldots v^{(i)}_N},
\end{eqnarray}
with $\mu_i\geq 0$. Since $\rho$ is symmetric one has $\ket{v^{(i)}_1\ldots v^{(i)}_N}\in\cals$. The symmetry imposes that $\ket{v^{(i)}_1}=\ldots=\ket{v^{(i)}_N}$. As spin-$\frac12$ states are all coherent states and from Eq.~\eqref{alphajalpha} the tensor product of identical spin-$\frac12$ coherent states yields a spin-$j$ coherent states, this completes the proof. The converse is obvious, since inserting \eqref{alphajalpha} into \eqref{rhoascoh} shows that any classical state is separable and symmetric.
\end{proof}

Seeing a spin-$j$ state as a multipartite state allows one to define
partial operations on subsystems, such as partial tracing or partial transposition. An
important property of the tensor representation \eqref{rhoarbitrary}
is the following: the partial trace of a state $\rho$ with tensor
$X_{\mu_1 \mu_2 \ldots \mu_N}$, taken over $N-k$ qubits, is a symmetric
$k$-qubit state with tensor coefficients
$X_{\mu_1 \ldots\mu_k0\ldots0}$ \cite{PRL2015}. This will allow us to reexpress various separability criteria in terms of the $X_{\mu_1\mu_2\ldots\mu_N}$.

\subsection{Separability criteria}
\label{secPT}
Using the correspondence outlined above, classicality criteria can be
obtained from known separability criteria, such as the PPT criterion. Let us consider a bipartite quantum state
$\rho\in \cH_1\otimes\cH_2$, with $\cH_1,\cH_2$ two finite-dimensional
Hilbert spaces of dimension $d_1$ and $d_2$, respectively. The partial transpose of
$\rho$ with respect to subsystem $2$ is defined by 
\begin{equation}
\label{rhopt}
(\rho^{\pt})_{i_1i_2,j_1j_2}=\rho_{i_1j_2,j_1i_2},\quad 0\leq i_k,j_k\leq d_k-1.
\end{equation}
Peres \cite{Peres96} showed that positivity of the partial transpose matrix $\rho^{\pt}$ is a necessary condition for separability. It was conjectured \cite{Peres96} and later proved \cite{Horo3} that PPT is a necessary and sufficient condition in the case where $d_A=2$ and $d_B=2$ or $3$.

In the case of classical spin-$j$ states seen as fully separable
symmetric states of $N=2j$ spins-$\frac12$, PPT yields a necessary criterion
for any bipartition of the $N$ qubits into $r$ and $N-r$ qubits. As
the state is symmetric this criterion only depends on the number $r$
of the qubits and not on which qubits are chosen. We denote by PT$(N-r:r)$ the partial transpose matrix
associated with such a bipartition, where transposition only affects
the Hilbert space associated with the last $r$ qubits. For instance
for a 5-qubit separable state $\rho_1 \otimes\rho_2 \otimes\rho_3
\otimes\rho_4\otimes\rho_5$ we have PT$(3:2)=\rho_1 \otimes\rho_2
\otimes\rho_3 \otimes\rho_4^T\otimes\rho_5^T$. As  PT$(r:N-r)$ is the transpose of the matrix  PT$(N-r:r)$ we shall only consider the case $r\leq j$.

The Peres separability criterion \cite{Peres96} gives as a necessary classicality criterion
\begin{equation}
\label{critP}
\ppt(N-r:r)\geq 0,\quad 1\leq r\leq N/2.
\end{equation}
For states of 2 or 3 qubits, the Peres-Horodecki criterion \cite{Horo3} yields a necessary and sufficient separability condition that reads
\begin{equation}
\label{critPH}
\ppt(N-1:1)\geq 0.
\end{equation}
Equivalently, with $j=N/2$, this gives a necessary and sufficient classicality condition for spin-$j$ states with $j=1$ or $j=3/2$.

\section{PPT and tensor representation}
\label{evenPPT}

\subsection{Matrix $\matrixT$ for equal bipartition}
In this section we reformulate the classicality criterion \eqref{critP} for integer $j$ and equal bipartition $(j:j)$ in terms of tensor entries $X_{\mu_1\mu_2\ldots\mu_N}$. We start by introducing the $4^j\times 4^j$ matrix
\begin{equation}
\label{etaGeneral}
\matrixT_{\bfmu,\bfnu}= X_{\mu_1\dots\mu_{j}\nu_1\ldots\nu_j},
\end{equation}
where matrix indices are vectors $\bfmu=(\mu_{1}\dots\mu_j)$ and
$\bfnu=(\nu_{1}\dots\nu_{j})$, $0\leq\mu_i,\nu_i\leq 3$. (In this paper we use commas to
separate the two (multi-)indices of a square matrix, while tensor
indices have no commas.) According to the definition of
$X_{\mu_1\dots\mu_{N}}$, the matrix elements of $\matrixT$ can all be
obtained as expectation values of tensor products of Pauli
operators. The matrix $\matrixT$ is real and symmetric. It turns out,
as we will show, that $\rho^{\pt}=\ppt(j:j)$ is similar to a multiple
of $\matrixT$, that is, there exists a unitary matrix $R$ and a (positive)
constant $\lambda$ such that $R^\dagger \rho^{\pt} R = \lambda
\matrixT$. In particular this implies that for the equal bipartition
$(j:j)$, the positivity of the partial transpose $\rho^{\pt}$ is
equivalent to the positivity of the matrix $\matrixT$, so that the
corresponding necessary classicality criterion can be expressed as
$\matrixT\geq 0$. We first examine the cases of small $j$ and then
move on to the general situation.

\subsection{Spin-1 case}
In the spin-1 case the matrix $\matrixT$ in \eqref{etaGeneral} coincides with the $4\times 4$ matrix $X$, since the multi-indices $\bfmu$ and $\bfnu$ reduce to single indices $\mu$ and $\nu$, $0\leq\mu,\nu\leq 3$. Let $\rho^{\pt}$ be the partial transpose of the spin-1 state $\rho$ written in the canonical basis of two qubits; it can be expressed as in \eqref{rhopt}. We want to find a $4\times 4$ unitary matrix $R$ with the property that
\begin{equation}
\label{regrouping}
(R^\dagger)_{\mu,{i_1}{i_2}} \rho_{i_1j_2,j_1i_2} R_{{j_1}{j_2},\nu} =\lambda X_{\mu,\nu}
\end{equation}
with $0\leq i_1,i_2,j_1,j_2\leq 1$ and $0\leq\mu,\nu\leq 3$. 
Suppose that $\rho$ is a coherent state. Then the left-hand side of Eq.~\eqref{regrouping} reads
\begin{equation}
 \label{regroupingb}
(R^\dagger)_{\mu,i_1i_2} (\ket{\alpha}\bra{\alpha})_{i_1, j_1} \,
(\ket{\alpha}\bra{\alpha})_{j_2,i_2} R_{j_1j_2,\nu}, 
\end{equation}
which can be rewritten as 
\begin{equation}
\label{regroupingc}
\bra{\alpha}_{{i_2}} (R^\dagger)_{\mu,{i_1}{i_2}} \ket{\alpha}_{{i_1}} \,\,   \bra{\alpha}_{{j_1}} R_{{j_1}{j_2},\nu} \ket{\alpha}_{{j_2}},
\end{equation}
while from Eq.~\eqref{coorcoh} the tensor coordinates of $\rho$ can be expressed as $X_{\mu,\nu}=\bra{\alpha}\sigma^\mu \ket{\alpha}\bra{\alpha}\sigma^\nu \ket{\alpha}$. One easily checks that a possible choice of $R$ that complies with Eq.~\eqref{regrouping} is
\begin{equation}
\label{Rspin1}
R_{{i_1}{i_2},\mu}=\frac{1}{\sqrt{2}}\sigma^\mu_{i_1,i_2}
\end{equation}
together with $\lambda=1/2$. Since $R$ and $\lambda$ chosen are both independent of $\ket{\alpha}$, they will in fact fulfill Eq.~\eqref{regrouping} for any coherent state. As any density matrix $\rho$ can be expanded as a linear combination of coherent states (as in \eqref{rhoascoh}, but possibly with a negative $P$-function), $R$ and $\lambda$ will be suited for any density matrix. Moreover, the matrix $R$ is unitary since
\begin{equation}
\label{RRspin1}
(R^\dagger R)_{\mu,\nu}=\frac12\tr\{\sigma^{\mu}\sigma^{\nu}\}=\delta_{\mu,\nu},
\end{equation}
with $\delta_{\mu,\nu}$ the Kronecker symbol. Thus, $(R^\dagger \rho^{\pt} R)_{\mu,\nu} = \lambda X_{\mu,\nu}$, so that the PPT criterion PT$(1:1)\geq 0$ is equivalent to positivity of the $4\times 4$ matrix $(X_{\mu,\nu})_{0\leq \mu,\nu\leq 3}$.

\subsection{Spin-2 case}
For spin 2, the matrix $\matrixT$ is indexed by multi-indices $(\mu_1\mu_2)$ and $(\nu_1\nu_2)$, while the matrix $\rho$ expressed in the computational basis of qubits is indexed by multi-indices $(i_1i_2i_3i_4)$ and $(j_1j_2j_3j_4)$, with again $i_k,j_k\in \{0,1\}$ and $\mu_k,\nu_k \in \{0,1,2,3 \}$. We are looking for a unitary matrix $R$ and a constant $\lambda$ such that $R^\dagger \rho^{\pt} R = \lambda \matrixT$, with $\rho^{\pt}$ the partial transpose taken over the last two qubits. As before, we can first consider the case where $\rho$ is a coherent state. Explicitly, the analog of Eq.~\eqref{regroupingb} for the components of $R^\dagger \rho^{\pt} R$ reads
\begin{align}
(R^\dagger)_{\mu_1\mu_2,{i_1}{i_2}{i_3}{i_4}} \ketbra{\alpha}_{{i_1},{j_1}} \,\, \ketbra{\alpha}_{{i_2},{j_2}}\hfill \nonumber\\
\,\,\times \ketbra{\alpha}_{{j_3},{i_3}} \,\, \ketbra{\alpha}_{{j_4},{i_4}} R_{{j_1}{j_2}{j_3}{j_4},\nu_1\nu_2},
\end{align}
while the analog of \eqref{regroupingc} reads
\begin{align}
\bra{\alpha}_{{i_3}} \bra{\alpha}_{{i_4}} (R^\dagger)_{\mu_1\mu_2,{i_1}{i_2}{i_3}{i_4}} \ket{\alpha}_{{i_1}}\ket{\alpha}_{{i_2}}\nonumber \\
\,\,\times  \bra{\alpha}_{{j_1}}\bra{\alpha}_{{j_2}} R_{{j_1}{j_2}{j_3}{j_4},\nu_1\nu_2} \ket{\alpha}_{{j_3}}\ket{\alpha}_{{j_4}}.
\label{eqq1}
\end{align}
The matrix $\matrixT$ can now be written as
\begin{equation}
\label{eqq2}
\bra{\alpha}\sigma^{\mu_1} \ket{\alpha} \bra{\alpha}\sigma^{\mu_2} \ket{\alpha}  \bra{\alpha}\sigma^{\nu_1} \ket{\alpha}\bra{\alpha}\sigma^{\nu_2} \ket{\alpha}.
\end{equation}
A choice of $R$ and $\lambda$ that fulfills the required relation between \eqref{eqq1} and \eqref{eqq2} is
\begin{equation}
\label{Rspin2}
R_{{i_1}{i_2}{i_3}{i_4},\mu_1\mu_2}=\frac12\sigma^{\mu_1}_{{i_1},{i_3}}\sigma^{\mu_2}_{{i_2},{i_4}}.
\end{equation}
The corresponding value of $\lambda$ is then $\lambda=1/4$. Note that other choices are possible for $R$: a different solution would be $\frac12\sigma^{\mu_1}_{{i_1},{i_4}}\sigma^{\mu_2}_{{i_2},{i_3}}$. Since $R$ and $\lambda$ are independent of $\ket{\alpha}$, they are valid for any coherent state and thus for any density matrix $\rho$. Unitarity of the matrix $R$ comes from the identity
\begin{equation}
(R^\dagger R)_{\bfmu,\bfnu}=\frac14\tr\{\sigma^{\mu_1}\sigma^{\nu_1}\}\tr\{\sigma^{\mu_2}\sigma^{\nu_2}\}=\delta_{\mu_1\nu_1}\delta_{\mu_2\nu_2},
\end{equation}
with $\bfmu=(\mu_1\mu_2)$, $\bfnu=(\nu_1\nu_2)$. Therefore, the necessary  PPT criterion \eqref{critP} for spin-2 states can be expressed as $\matrixT\geq 0$, where $\matrixT$ is the $16\times 16$ matrix defined by $\matrixT_{\bfmu,\bfnu}=X_{\mu_1\mu_2\nu_1\nu_{2}}$.

\subsection{General case}
The above construction easily generalises to higher integer spin sizes. For spin $j$ the $4^j\times 4^j$ matrix $R$ reads
\begin{equation}
\label{Rgeneral}
R_{\bfi,\bfmu}=\frac{1}{2^{j/2}}\prod_{k=1}^j\sigma^{\mu_k}_{i_k,i_{k+j}},
\end{equation}
where $\bfi=(i_1 i_2 \ldots i_N)$ and $\bfmu=(\mu_1\mu_2\ldots\mu_j)$, with $0 \leq \mu_k\leq 3$ and $0 \leq i_k\leq 1$. Note that each Pauli matrix is indexed by one index associated with an non-transposed qubit and one associated with a transposed qubit. Any such pairing would yield a valid $R$. It is easy to check that matrices $R$ are unitary and such that $R^\dagger \rho^{\pt} R = \lambda \matrixT$, with $\rho^{\pt}=\ppt(j:j)$ and $\lambda=1/2^j$. Thus, the corresponding PPT criterion yields the classicality criterion $\matrixT\geq 0$.

\section{PPT for any bipartition}
\label{generalPPT}

\subsection{$\matrixTr$ matrices}
The results of Section \ref{evenPPT} can be further generalized to uneven bipartitions of symmetric states. In this section we show that matrices $\ppt(N-r:r)$ are similar to a multiple of matrices $\matrixTr$ defined by
\begin{equation}
\label{etarGeneral}
\matrixTr_{\bfmu\;\bfi, \bfnu\;\bfi'}=X_{\tau_1\ldots\tau_{N-2r}\mu_1\ldots \mu_r \nu_1 \ldots \nu_{r}} \prod_{k=1}^{N-2r}\sigma^{\tau_{k}}_{i_k,i'_k}
\end{equation}
where $\bfmu=(\mu_{1}\dots\mu_r)$, $\bfnu=(\nu_{1}\dots\nu_{r})$, $\bfi=(i_1\ldots i_{N-2r})$ and $\bfi'=(i'_1 \dots i'_{N-2r})$ are multi-indices with $0 \leq \mu_k\leq 3$ and $0 \leq i_k,i'_k \leq 1$, and summation over the $\tau_k\in \{0,1,2,3\}$ is implicit. In this definition, indices $\bfnu$ are associated with the transposed subspace, while indices $\bftau$ and $\bfmu$ are associated with the non-transposed one. Matrices $\matrixTr$ are of size $4^j\times 4^j$. In the case of equal bipartition $r=j$, Eq.~\eqref{etarGeneral} reduces to Eq.~\eqref{etaGeneral}.

\subsection{Spin 3/2}
Let us start by considering the smallest-size case. Let $\rho$ be a spin-$3/2$ state and $\rho^{\pt}=\ppt(2:1)$, its transpose with respect to the third qubit. The matrix $\matrixT^{(1)}$ in Eq.~\eqref{etarGeneral} is given by
\begin{equation}
\label{etar32}
\matrixT^{(1)}_{\mu\;i,\nu\;i'}=X_{\tau\mu\nu}\sigma^\tau_{i,i'}. 
\end{equation}
Building on the results of the previous section, it is easy to construct a unitary matrix $R$ such that  $R^\dagger \rho^{\pt} R=\lambda\matrixTr$. As before we consider the case where $\rho$ is a coherent state. In such a case, $R^\dagger \rho^{\pt} R$ reduces to
\begin{align}
(R^\dagger)_{\mu\;i,a_1 a_2 a_3} \ketbra{\alpha}_{{a_1},{b_1}}\hspace{2cm}\nonumber\\
\times\ketbra{\alpha}_{{a_2},{b_2}}\ketbra{\alpha}_{{b_3},{a_3}} R_{{b_1}{b_2}{b_3},\nu\; i'},
\end{align}
with $0\leq a_k,b_k\leq 1$, or equivalently
\begin{align}
\bra{\alpha}_{a_3}(R^\dagger)_{\mu\; i,a_1a_2a_3} \ket{\alpha}_{a_1}\ket{\alpha}_{a_2}
\hspace{1cm}\nonumber\\ 
\times\bra{\alpha}_{b_1}\bra{\alpha}_{b_2} R_{{b_1}{b_2}{b_3},\nu\; i'}\ket{\alpha}_{b_3},
\label{spin32a}
\end{align}
while the matrix $\matrixT^{(1)}$  defined in \eqref{etar32} can be written for this coherent state $\ket{\alpha}$ as
\begin{equation}
\bra{\alpha}\sigma^{\mu} \ket{\alpha} \bra{\alpha}\sigma^{\nu} \ket{\alpha}(2\ketbra{\alpha})_{i,i'}
\label{spin32b}
\end{equation}
(we used the fact that $\frac12 n_{\tau}\sigma^{\tau}=\ketbra{\alpha}$). Identifying Eqs.~\eqref{spin32a} and \eqref{spin32b} up to a constant we see that a unitary $R$ can be defined for instance as 
\begin{equation}
\label{R32}
R_{a_1a_2a_3,\mu\; i}=\frac{1}{\sqrt{2}}\delta_{a_1,i}\sigma^\mu_{a_2,a_3}.
\end{equation}
In fact, the indices $a_2, a_3$ of the matrix $\sigma^\mu$ in \eqref{R32} have to pair any index associated with the non-transposed subspace with an index associated with the transposed subspace, while the delta function pairs the remaining indices in \eqref{spin32a}, leading to the projector $\ketbra{\alpha}$ in \eqref{spin32b}. Unitarity of $R$ is easily verified, since
\begin{equation}
\label{R32unitary}
(R^{\dagger}R)_{\mu\; i,\nu\; i'}=\frac{1}{2}\tr\{\sigma^\mu\sigma^\nu\}\sum_{i_2}\delta_{i,i_2}\delta_{i_2,i'}=\delta_{\mu,\nu}\delta_{i,i'}.
\end{equation}
As in the equal bipartition case, linearity ensures that $R$ defined in \eqref{R32} together with $\lambda=1/4$ is such that $R^\dagger \rho^{\pt} R=\lambda\matrixTr$.

\subsection{General $R$ matrices}
The above case contains the essence of the general proof and generalises to arbitrary values of $j$ and $r$. In order to recover matrix $\matrixTr$ from PT$(N-r:r)$, we have to construct a matrix $R$ built out of products of $\sigma^\mu$ matrices and Kronecker deltas, such that the Pauli matrices pair $r$ indices among those associated with the non-transposed subspace together with all $r$ indices associated with the transposed subspace. The remaining $N-2r$ indices, corresponding to the remaining part of the non-transposed subspace, go into Kronecker deltas. More precisely, we choose these latter indices to be the $N-2r$ first ones, and to pair indices $k$ with $k+r$ for $N-2r+1\leq k \leq N-r$. We thus define matrices $\Rr$ by
\begin{align}
\label{Rmat}
\Rr_{\bfa, \bfmu\; \bfi }=\frac{1}{2^{r/2}}\prod_{k=1}^{N-2r}\delta_{a_k,i_k}
\prod_{k=1}^r\sigma^{\mu_k}_{a_{N-2r+k},a_{N-r+k}}
,
\end{align}
with $\bfa=(a_1 \dots a_N)$, $\bfmu=(\mu_1 \dots \mu_{r})$ and $\bfi=(i_1 \dots i_{N-2r})$, with $\mu_k \in \{0,1,2,3 \}$ and $a_k, i_k\in \{0,1 \}$. One can check, as above, that $\Rr$ are unitary and such that 
\begin{equation}
\label{RrhoRgeneral}
(\Rr)^{\dagger}\rho^{\pt}\Rr=\frac{1}{2^{N-r}}\matrixTr
\end{equation}
with $\rho^{\pt}=\ppt(N-r:r)$. Unitarity trivially comes from the fact that indices of the Pauli matrices and the Kronecker deltas in \eqref{Rmat} are all distinct, so that the identity \eqref{RRspin1} can be applied to each pair of matrices. To show \eqref{RrhoRgeneral}, we first write $\rho^{\pt}$ in the computational basis with the help of the tensor representation. As explained in Section \ref{sec:classical}, the expansion \eqref{rhoarbitrary} can be obtained by projecting tensor products of Pauli matrices onto the symmetric subspace. In the computational basis of $N$ qubits, $\rho$ can thus be expressed as 
\begin{equation}
\rho=\frac{1}{2^N}\,X_{\mu_1\mu_2\ldots\mu_N} \sigma^{\mu_1}\otimes\sigma^{\mu_2}\otimes\ldots\otimes\sigma^{\mu_{N}},
\end{equation}
so that $\rho^{\pt}$ reads
\begin{equation}
\label{rhoPTcompbasis}
\rho^{\pt}_{\bfa,\bfb}=\frac{1}{2^{N}} X_{\tau_1 \cdots \tau_{N}} \prod_{k=1}^{N-r}\sigma^{\tau_k}_{a_k,b_k}\prod_{k=N-r+1}^{N}\sigma^{\tau_k}_{b_k,a_k}
\end{equation}
with $\bfa=(a_1 \dots a_N)$ and $\bfb=(b_1 \dots b_N)$, $a_k,b_k\in\{0,1\}$. The left-hand side of \eqref{RrhoRgeneral} has components
\begin{equation}
((\Rr)^{\dagger}\rho^{\pt}\Rr)_{\bfmu\; \bfi , \bfnu\; \bfi'}=(\Rr_{\bfa,\bfmu\; \bfi})^*\rho^{\pt}_{\bfa,\bfb}\Rr_{\bfb,\bfnu\; \bfi'}
\end{equation}
where $*$ denotes complex conjugation. Using \eqref{Rmat} and \eqref{rhoPTcompbasis}, this can be expressed as
\begin{align}
\label{bigproduct}
\frac{1}{2^{r+N}}X_{\tau_1 \cdots \tau_{N}}\prod_{k=1}^{N-2r}\delta_{a_k,i_k}
\prod_{k=N-2r+1}^{N-r}\sigma^{\mu_{k-N+2r}}_{a_{k+r},a_k}\prod_{k=1}^{N-r}\sigma^{\tau_k}_{a_k,b_k}
\nonumber\\
\times\prod_{k=N-r+1}^{N}\sigma^{\tau_k}_{b_k,a_k}
\prod_{k=1}^{N-2r}\delta_{b_k,i'_k}\prod_{k=N-2r+1}^{N-r}\sigma^{\nu_{k-N+2r}}_{b_k,b_{k+r}}.
\end{align}
The above product contains terms
\begin{equation}
\label{terms1}
\delta_{a_k,i_k}\sigma^{\tau_k}_{a_k,b_k}\delta_{b_k,i'_k}=\sigma^{\tau_k}_{i_k,i'_k}
\end{equation}
for $1\leq k \leq N-2r$, terms
\begin{equation}
\label{terms2}
\sigma^{\mu_{k-N+2r}}_{a_{k+r},a_k}\sigma^{\tau_k}_{a_k,b_k}\sigma^{\nu_{k-N+2r}}_{b_k,b_{k+r}}
=(\sigma^{\mu_{k-N+2r}}\sigma^{\tau_k}\sigma^{\nu_{k-N+2r}})_{a_{k+r},b_{k+r}}
\end{equation}
for $N-2r+1\leq k \leq N-r$, and terms 
\begin{equation}
\label{terms3}
\sigma^{\tau_k}_{b_k,a_k}
\end{equation}
for $N-r+1\leq k \leq N$ (recall that we are considering a case where $N-r\geq r$). Taking the product of all terms \eqref{terms1}--\eqref{terms3} and summing over the remaining $a_k$ and $b_k$ (those with $N-r+1\leq k \leq N$), \eqref{bigproduct} becomes
\begin{equation}
\label{prodtr}
\frac{X_{\tau_1 \cdots \tau_{N}}}{2^{r+N}}\prod_{k=1}^{N-2r}\sigma^{\tau_k}_{i_k,i'_k}\prod_{k=1}^r\tr\{\sigma^{\mu_k}\sigma^{\tau_{k+N-2r}}\sigma^{\nu_k}\sigma^{\tau_{k+N-r}}\}.
\end{equation}
As can be checked explicitly, one has the identity
\begin{equation}
\label{identity_trace}
\frac14 y_{\tau, \tau'}\tr\{\sigma^{\mu}\sigma^{\tau}\sigma^{\nu}\sigma^{\tau'}\}=y_{\mu, \nu}
\end{equation}
for any real symmetric matrix $(y_{\mu,\nu})_{0\leq\mu,\nu\leq 3}$ such that $\sum_{a=1}^3y_{aa}=y_{00}$. Applying this identity to the summation over pairs of indices $(\tau_{k+N-2r},\tau_{k+N-r})$ for $1\leq k\leq r$ in \eqref{prodtr} (and using property \eqref{propx} of the tensor), we recover the term $X_{\tau_1 \ldots \tau_{N-2r} \mu_1 \ldots \mu_r \nu_1 \ldots \nu_{r}}$ of \eqref{etarGeneral}. The product of terms \eqref{terms1} yields the Pauli matrix terms in \eqref{etarGeneral}. The overall remaining factor is $\lambda=1/2^{N-r}$. This proves Eq.~\eqref{RrhoRgeneral}.

\vspace{2cm}

\section{Some consequences}
\label{applications}

\subsection{PPT criteria}
As mentioned in Section \ref{secPT}, the PPT separability criterion provides necessary, and in some instances sufficient, classicality criteria. The previous sections have shown that the partial transpose takes a very simple form for symmetric states expressed as in \eqref{rhoarbitrary}. Thus each PPT criterion is equivalent to a linear matrix inequality $\matrixTr\geq 0$. In the simplest case of spin-1 states, $\matrixTr$ is given by Eq.~\eqref{etaGeneral}, so that the PPT criterion PT$(1:1)\geq 0$ is equivalent to $X\geq 0$ for the $4\times 4$ matrix $(X_{\mu,\nu})_{0\leq \mu,\nu\leq 3}$. This was already observed in \cite{fabian1}, where the same relation between $\rho^\pt$ and $X$ was obtained. Our present results generalize this relation: For integer spin and equal bipartition $r=j$, the PT$(j:j)$ criterion is expressed in a very transparent way in our tensor language, as the positivity of the matrix $(T_{\bfmu,\bfnu})$ indexed by $j$-tuples of indices and defined in \eqref{etaGeneral}. More generally, each PPT criterion yields a classicality criterion as the positivity of a matrix $T^{(r)}$.

In the case of spin-3/2, using the results of Section \ref{generalPPT}, a necessary and sufficient classicality criterion can be expressed as $\matrixT^{(1)}\geq 0$, where $\matrixT^{(1)}$ is defined in \eqref{etar32}. In terms of the tensor entries, this criterion reads
\begin{widetext}
\begin{small}
\begin{equation}
\label{CNS32}
\left(
\begin{array}{cccccccc}
 X_{000}+X_{003} & X_{001}-\mathrm{i} X_{002} & X_{001}+X_{013} &
   X_{011}-\mathrm{i} X_{012} & X_{002}+X_{023} & X_{012}-\mathrm{i} X_{022} &
   X_{003}+X_{033} & X_{013}-\mathrm{i} X_{023} \\
 X_{001}+\mathrm{i} X_{002} & X_{000}-X_{003} & X_{011}+\mathrm{i} X_{012} &
   X_{001}-X_{013} & X_{012}+\mathrm{i} X_{022} & X_{002}-X_{023} &
   X_{013}+\mathrm{i} X_{023} & X_{003}-X_{033} \\
 X_{001}+X_{013} & X_{011}-\mathrm{i} X_{012} & X_{011}+X_{113} &
   X_{111}-\mathrm{i} X_{112} & X_{012}+X_{123} & X_{112}-\mathrm{i} X_{122} &
   X_{013}+X_{133} & X_{113}-\mathrm{i} X_{123} \\
 X_{011}+\mathrm{i} X_{012} & X_{001}-X_{013} & X_{111}+\mathrm{i} X_{112} &
   X_{011}-X_{113} & X_{112}+\mathrm{i} X_{122} & X_{012}-X_{123} &
   X_{113}+\mathrm{i} X_{123} & X_{013}-X_{133} \\
 X_{002}+X_{023} & X_{012}-\mathrm{i} X_{022} & X_{012}+X_{123} &
   X_{112}-\mathrm{i} X_{122} & X_{022}+X_{223} & X_{122}-\mathrm{i} X_{222} &
   X_{023}+X_{233} & X_{123}-\mathrm{i} X_{223} \\
 X_{012}+\mathrm{i} X_{022} & X_{002}-X_{023} & X_{112}+\mathrm{i} X_{122} &
   X_{012}-X_{123} & X_{122}+\mathrm{i} X_{222} & X_{022}-X_{223} &
   X_{123}+\mathrm{i} X_{223} & X_{023}-X_{233} \\
 X_{003}+X_{033} & X_{013}-\mathrm{i} X_{023} & X_{013}+X_{133} &
   X_{113}-\mathrm{i} X_{123} & X_{023}+X_{233} & X_{123}-\mathrm{i} X_{223} &
   X_{033}+X_{333} & X_{133}-\mathrm{i} X_{233} \\
 X_{013}+\mathrm{i} X_{023} & X_{003}-X_{033} & X_{113}+\mathrm{i} X_{123} &
   X_{013}-X_{133} & X_{123}+\mathrm{i} X_{223} & X_{023}-X_{233} &
   X_{133}+\mathrm{i} X_{233} & X_{033}-X_{333} \\
\end{array}
\right)\geq 0.
\end{equation}
\end{small}
\end{widetext}
This matrix inequality can in turn be expressed as positivity of a $16\times 16$ real symmetric matrix, whose entries are of the form $\pm X_{\mu_1\mu_2\mu_3}$, i.~e.~, $\pm\langle \sigma^{\mu_1}\otimes\sigma^{\mu_2}\otimes\sigma^{\mu_{3}}\rangle$, which provides a necessary and sufficient classicality condition as positivity of a matrix of observables.

\subsection{Correlation matrices}
Let $X$ be the tensor representation \eqref{coeffdecompPauli} of a
spin-$j$ state $\rho$ with $j$ integer. We define correlation matrices
associated with the tensor $X$ as the $4^r\times 4^r$ matrices ($1\le
r\le j$)
\begin{equation}
\label{Cmatrix}
C^{(r)}_{\bfmu_r,\bfnu_r}=X_{\bfmu_r \bfnu_r \bfz_{N-2r}}-X_{\bfmu_r \bfz_{N-r}}X_{\bfnu_r \bfz_{N-r}}, 
\end{equation}
where $\bfmu_r=(\mu_1 \dots \mu_{r})$, $\bfnu_r=(\nu_1 \dots \nu_{r})$, and $\bfz_k$ is the zero vector of length $k$. Since the first line and column of $C^{(r)}$ are indexed by $\bfz_r$, and $X_{0\ldots 0}=1$, it takes the form
\begin{equation}
C^{(r)}=\left(
\begin{array}{cccc}
0 & 0 & \cdots & 0\\
0 & & &    \\
\vdots & & S^{(r)} &    \\
0 & & &    \\
\end{array}
\right),
\end{equation}
where the matrix $S^{(r)}$ is of size $(4^r-1)\times(4^r-1)$. In terms of the entries of the matrix $\matrixT$ defined in \eqref{etaGeneral}, $S^{(r)}$ can be expressed as 
\begin{equation}
\label{Smatrix}
S^{(r)}_{\bfmu_r,\bfnu_r}=\matrixT_{\bfmu_r \bfz_{j-r},\bfnu_r \bfz_{j-r}}-\matrixT_{\bfmu_r \bfz_{j-r},\bfz_{j}}\matrixT_{\bfnu_r \bfz_{j-r},\bfz_{j}}.
\end{equation}
The matrix $S^{(r)}$ can thus be interpreted as the Schur complement
of the matrix
$(\matrixT_{\bfmu_r \bfz_{j-r},\bfnu_r \bfz_{j-r}})_{\bfmu_r,\bfnu_r}$ with respect to the upper left entry $\matrixT_{\bfz_j,\bfz_j}=1$. The matrix $(\matrixT_{\bfmu_r\,\bfz_{j-r},\bfnu_r\,\bfz_{j-r}})_{\bfmu_r,\bfnu_r}$ is the restriction of $\matrixT$ to its $4^r$ first lines and columns. This $4^r\times 4^r$ subblock coincides with the matrix $\matrixT$ associated with
the spin-$r$ state $\rho_r$ obtained from $\rho$ by tracing out $N-2r$
qubits. Since positivity of a matrix is equivalent to positivity of
its Schur complement (if the part complemented is itself positive),
one has that the upper left $4^r\times 4^r$ block of $\matrixT$ is positive iff $C^{(r)}\geq 0$. Together with the results of the previous sections, this shows that the PPT criterion PT$(j:j)\geq 0$ applied to $\rho$ is equivalent to positivity of the correlation matrix $C^{(j)}$, and more generally the PPT criterion PT$(r:r)\geq 0$ applied to the reduced density matrix $\rho_r$ is equivalent to positivity of the correlation matrix $C^{(r)}$. 

If $\rho$ is a classical state, then all its reduced density matrices $\rho_r$ are classical as well. The PPT criterion thus leads to a sequence of necessary classicality conditions $C^{(r)}\geq 0$. These conditions are those obtained by different means in \cite{UshPraRaj07}, where the so-called 'intergroup covariance matrices' coincide with our matrices $C^{(r)}$. This also allows us to recover results from \cite{TothGuehne} that the partial transpose criterion for partition into two equally sized subsystems is equivalent to positivity of the correlation matrix of local orthogonal observables.

From the above considerations, we see that all these necessary conditions are encompassed in a compact way in the single condition $\matrixT\geq 0$. This latter condition is not sufficient, nor is the condition that all partial transposes be positive. For instance, there exist symmetric 4-qubit entangled states for which all partial transposes are positive \cite{TurLew12}.

\subsection{Teleportation and generalized magic bases} 
\label{magicbases}

The matrix $R_{\bfi,\mu}=\frac{1}{\sqrt{2}}\sigma^\mu_{i_1,i_2}$ with $\bfi=(i_1,i_2)$ defined in \eqref{Rspin1} can be written out explicitly in the computational basis as
\begin{equation}
\label{R22}
R=\frac{1}{\sqrt{2}}\left(
\begin{array}{cccc}
 1 & 0 & 0 & 1 \\
 0 & 1 & -\mathrm{i} & 0 \\
 0 & 1 & \mathrm{i} & 0 \\
 1 & 0 & 0 & -1 \\
\end{array}
\right).
\end{equation}
The $\mu$th column of $R$ contains the elements of the Pauli matrix $\sigma^\mu$ (up to normalization). These are equal, up to a phase factor, to the two-qubit Bell states. More precisely, the Bell states are the columns of the matrix $\tilde{R}_{{i_1}{i_2},\mu}=\frac{1}{\sqrt{2}}\tilde{\sigma}^{\mu}_{i_1,i_2}$, where $\tilde{\sigma}^{\mu}=\sigma^{\mu}$ for $\mu\neq 2$ and $\tilde{\sigma}^{\mu}=$i$\sigma^{\mu}$ for $\mu=2$. They are also proportional to the magic basis introduced in \cite{HilWoo97}: namely, the three last columns of $R$ have to be multiplied by $-i$ in order to recover the magic basis of \cite{HilWoo97}. We recall that, among other properties, the magic basis is such that when a state $\ket{\psi}$ is written in this basis, with some coefficients $\alpha_i,1\leq i \leq 4$, then its concurrence is given by $C(\ket{\psi})=|\sum_{i=1}^{4}\alpha_i^2|$.

Bell states are used in the quantum teleportation protocol of a single
qubit \cite{BenBraCre93}. If Alice and Bob share a Bell state, it is possible for them to teleport a one-qubit state by exchanging only two classical bits. In a similar spirit, four-qubit generalized Bell states $\ket{g_i}$, $1\leq i\leq 16$, were introduced in \cite{Rig05}: if Alice and Bob share one of these generalized Bell state, they are able to teleport a two-qubit pure state by exchanging four classical bits (the protocol of \cite{Rig05} is essentially the same as in the one-qubit case). It turns out that the columns of our spin-$2$ matrix $R$, defined explicitly in \eqref{Rspin2}, are equal, up to a phase factor, to the 16 states $\ket{g_i}$. More precisely, the $\ket{g_i}$ of \cite{Rig05} are exactly the columns of the matrix  
\begin{equation}
\label{Rspin2bis}
\tilde{R}_{{i_1}{i_2}{i_3}{i_4},\mu_1\mu_2}=\frac12\tilde{\sigma}^{\mu_1}_{{i_1},{i_3}}\tilde{\sigma}^{\mu_2}_{{i_2},{i_4}}
\end{equation}
(again the $\tilde{\sigma}$ are such that $\tilde{\sigma}^\mu=$i$\sigma^\mu$ for $\mu=2$, and $\sigma^\mu$ otherwise). The generalized Bell basis also provides a generalization of the magic basis to higher qubits. The 2-qubit magic basis $\ket{e_i}$, $1\leq i\leq 16$, in \cite{Rig05} is constructed by multiplying the $\ket{g_i}$ by appropriate phases. A state expressed in this basis as $\ket{\psi}=\sum_{i=1}^{16}\alpha_i\ket{e_i}$ is then such that the generalized concurrence \cite{WonChr01} is given by $C(\ket{\psi})=|\sum_{i=1}^{16}\alpha_i^2|$. We can recover the magic basis $\ket{e_i}$ just by multiplying by i the 8 columns of $\tilde{R}$ indexed by pairs $(\mu_1,\mu_2)$ such that $|\mu_1-\mu_2|=1$. Our formula thus provides a very compact form both for the Bell states appearing in the two-qubit teleportation protocol and for the generalized magic basis of \cite{Rig05}.

It is clear from Eq.~\eqref{Rspin2bis}, and from the general form \eqref{Rgeneral} of matrices $R$, that this approach can be straightforwardly generalized to an arbitrary number of qubits. The $N$-qubit teleportation protocol proposed in \cite{Rig05} was obtained from the action of products of the form $(\sigma^z)^{\alpha}(\sigma^x)^{\beta}$, with $\alpha,\beta\in\{0,1\}$, on a state $\sum_{j=0}^{N-1}\ket{j}\ket{j}$. Using the fact that $\sigma^z\sigma^x=i\sigma^y$, one can check that the generalized Bell basis coincides, up to phases, with the columns of our matrices. In particular, this means that $R$ can also be interpreted as the unitary matrix that Alice has to apply on her side to make a Bell measurement in the $N$-qubit teleportation protocol.

\section{Conclusion}
The present results provide a unifying framework for
various concepts dealing with symmetric states.
It allows us to reformulate
several known results 
in a much simpler way. In the language of the tensor representation,
criteria such as the PPT separability criterion can be expressed in a
much more transparent way by positivity of the matrix $\matrixT$. In
particular, this allows one 
to directly relate the partial transpose to correlations 
 of observables, which provides a new physical interpretation of the
 partial transpose beyond time-reversal. Note that the matrix $R$ in
 \eqref{R22} was used in \cite{BhoShuLak13} to generate local unitary
 invariants in terms of partial transpose and realignments. It
 may be possible to extend our expressions to that setting as well.\\

{\bf Acknowledgments:} We thank the Deutsch-Franz\"osische
Hochschule (Universit\'e franco-allemande) for support, grant
number CT-45-14-II/2015.


\begin{thebibliography}{99}
\bibitem{4Hor09} R.~Horodecki, P.~Horodecki, M.~Horodecki, and K.~Horodecki, Rev.~Mod.~Phys.~{\bf 81}, 865 (2009).

\bibitem{BenBraCre93} C.~H.~Bennett, G.~Brassard, C.~Cr\'epeau, R.~Jozsa, A.~Peres, and W.~K.~Wootters, Phys.~Rev.~Lett.~{\bf 70}, 1895 (1993).

\bibitem{Werner89} R.~F.~Werner, Phys.~Rev.~A {\bf 40}, 4277 (1989).

\bibitem{Peres96} A.~Peres, Phys.~Rev.~Lett.~{\bf 77}, 1413 (1996).

\bibitem{Horo3} M.~Horodecki, P.~Horodecki, and R.~Horodecki, Phys.~Lett.~A {\bf 223}, 1 (1996).

\bibitem{horodecki_separability_1997} P.~Horodecki, Phys.~Lett.~A {\bf 232}, 333 (1997).

\bibitem{amselem_experimental_2009} E.~Amselem, and M.~Bourennane, Nat.~Phys.~{\bf 5}, 748 (2009).

\bibitem{dur_classification_2000}  W.~D\"ur, and J.~I.~Cirac, Phys.~Rev.~A {\bf 61}, 042314 (2000).

\bibitem{dur_three_2000} W.~D\"ur, G.~Vidal, and J.~I.~Cirac, Phys.~Rev.~A {\bf 62}, 062314 (2001).

\bibitem{Acin2001} A.~Acin, D.~Bru{\ss}, M.~Lewenstein, and A.~Sanpera, Phys.~Rev.~Lett.~{\bf 87}, 040401 (2001).

\bibitem{gour_all_2010} G.~Gour and N.~R.~Wallach, J.~Math.~Phys.~{\bf 51}, 112201 (2010).

\bibitem{viehmann_polynomial_2011} O.~Viehmann, C.~Eltschka, and J.~Siewert, Phys.~Rev.~A {\bf 83}, 052330 (2011).

\bibitem{gour_classification_2013} G.~Gour and N.~R.~Wallach, Phys.~Rev.~Lett.~{\bf 111}, 060502 (2013).

\bibitem{TothGuehne} G.~T\'oth and O.~G\"uhne, Phys.~Rev.~Lett.~{\bf 102}, 170503 (2009).

\bibitem{bastin_operational_2009} T.~Bastin, S.~Krins, P.~Mathonet, M.~Godefroid, L.~Lamata, and E.~Solano, Phys.~Rev.~Lett.~{\bf 103}, 070503 (2009).

\bibitem{eckert_quantum_2002} K.~Eckert, J.~Schliemann, D.~Bru{\ss}, and M.~Lewenstein, Ann.~Phys.~{\bf 299}, 88 (2002).

\bibitem{TurLew12} J.~Tura, R.~Augusiak, P.~Hyllus, M.~Ku\'s, J.~Samsonowicz, and M.~Lewenstein, Phys.~Rev.~A {\bf 85}, 060302 (2012).

\bibitem{Toth09} G.~T\'oth and O.~G\"uhne, Phys.~Rev.~Lett.~{\bf 102}, 170503 (2009).

\bibitem{augusiak_entangled_2012} R.~Augusiak, J.~Tura, J.~Samsonowicz, and M.~Lewenstein, Phys.~Rev.~A {\bf 86}, 042316 (2012).

\bibitem{GirBraBra08} O.~Giraud, P.~Braun, and D.~Braun, Phys.~Rev.~A {\bf 78}, 042112 (2008).

\bibitem{Giraud10} O.~Giraud, P.~Braun, and D.~Braun, New J.~Phys.~{\bf 12}, 063005 (2010).

\bibitem{Martin10} J.~Martin, O.~Giraud, P.~A.~Braun, D.~Braun, and T.~Bastin, Phys.~Rev.~A {\bf 81}, 062347 (2010).

\bibitem{fabian1} F.~Bohnet-Waldraff, D.~Braun, and O.~Giraud, Phys.~Rev.~A {\bf 93}, 012104 (2016).

\bibitem{PRL2015} O.~Giraud, D.~Braun, D.~Baguette, T.~Bastin, and J.~Martin, Phys.~Rev.~Lett.~{\bf 114}, 080401 (2015).

\bibitem{wootters_entanglement_1998} W.~K.~Wootters, Phys.~Rev.~Lett.~{\bf 80}, 2245 (1998). 

\bibitem{UshPraRaj07} A.~R.~Usha Devi, R.~Prabhu, and A.~K.~Rajagopal, Phys.~Rev.~Lett.~{\bf 98}, 060501 (2007).

\bibitem{usha_devi_constraints_2007} A.~R.~Usha~Devi, M.~S.~Uma, R.~Prabhu, and A.~K.~Rajagopal, Phys.~Lett.~A {\bf 364}, 203 (2007).

\bibitem{KorCirLew05} J.~K.~Korbicz, J.~I.~Cirac, and M.~Lewenstein, Phys.~Rev.~Lett.~{\bf 95}, 120502 (2005).

\bibitem{UshRajKar13} A.~R.~Usha Devi, A.~K.~Rajagopal, Sudha, H.~S.~Karthik, and J.~Prabhu Tej, Quantum Inf.~Process.~{\bf 12}, 3717 (2013).

\bibitem{HilWoo97} S.~Hill and W.~K.~Wootters, Phys.~Rev.~Lett.~{\bf 78}, 5022 (1997). 

\bibitem{Rig05} G.~Rigolin, Phys.~Rev.~A {\bf 71}, 032303 (2005).


\bibitem{WonChr01} A.~Wong and N.~Christensen, Phys.~Rev.~A {\bf 63}, 044301 (2001).


\bibitem{BhoShuLak13} U.~T.~Bhosale, K.~V.~Shuddhodan, and A.~Lakshminarayan, Phys.~Rev.~A {\bf 87}, 052311 (2013).



\end{thebibliography}
\end{document}